\documentclass[pre,aps,superscriptaddress,twocolumn,amsmath,amssymb]{revtex4} 
\usepackage{graphics,placeins}
\usepackage{epsfig}
\usepackage{epstopdf}
\usepackage{amsmath,amssymb}
\usepackage{setspace}
\usepackage{graphicx}% Include figure files
\usepackage{dcolumn}% Align table columns on decimal point
\usepackage{bm}% bold math
\usepackage{color}
\usepackage{latexsym}
\usepackage{textcomp}
\usepackage{stackrel}
\usepackage{bbold}
\usepackage{xcolor}
\usepackage{xr-hyper}
\usepackage[citecolor=magenta,colorlinks=true]{hyperref}
\usepackage{lineno}
\usepackage{stackrel}
\usepackage[toc,page,titletoc]{appendix}
\usepackage[capitalise,nameinlink]{cleveref}

\begin{document}

\title{Emergent Elasticity in Amorphous Solids}

\author{Jishnu N. Nampoothiri}

\affiliation{Martin Fisher School of Physics, Brandeis University, Waltham, MA 02454 USA}
\affiliation{Centre for Interdisciplinary Sciences, Tata Institute of Fundamental Research, Hyderabad 500107, India}

\author{Yinqiao Wang}
\affiliation{Institute of Natural Sciences and School of Physics and Astronomy, Shanghai Jiao Tong University, Shanghai, 200240 China}

\author{Kabir Ramola}
\affiliation{Centre for Interdisciplinary Sciences, Tata Institute of Fundamental Research, Hyderabad 500107, India}

\author{Jie Zhang}
\affiliation{Institute of Natural Sciences and School of Physics and Astronomy, Shanghai Jiao Tong University, Shanghai, 200240 China}

\author{Subhro Bhattacharjee}
\affiliation{International Centre for Theoretical Sciences, Tata Institute of Fundamental Research, Bengaluru 560089, India}

\author{Bulbul Chakraborty}
\affiliation{Martin Fisher School of Physics, Brandeis University, Waltham, MA 02454 USA}
\email{bulbul@brandeis.edu}

\begin{abstract}
The mechanical response of naturally abundant amorphous solids such as gels, jammed grains, and biological tissues are not described by the conventional paradigm of broken symmetry that defines crystalline elasticity. In contrast, the response of such athermal solids are governed by local conditions of mechanical equilibrium, {\it i.e.}, force and torque balance of its constituents. Here we show that these constraints have the mathematical structure of a generalized electromagnetism, where the electrostatic limit successfully captures the anisotropic elasticity of amorphous solids. The emergence of elasticity from local mechanical constraints offers a new paradigm for systems with no broken symmetry, analogous to emergent gauge theories of quantum spin liquids. Specifically, our $U(1)$ rank-2 symmetric tensor gauge theory of elasticity translates to the electromagnetism of fractonic phases of matter with the stress mapped to electric displacement and forces to vector charges. We corroborate our theoretical results with numerical simulations of soft frictionless disks in both two and three dimensions, and experiments on frictional disks in two dimensions. We also present experimental evidence indicating that force chains in granular media are sub-dimensional excitations of amorphous elasticity similar to fractons.
\end{abstract}


\maketitle

\addcontentsline{toc}{section}{Introduction}
\paragraph*{Introduction:} Solids that emerge in strongly nonequilibrium processes such as jamming~\cite{Cates_1998,OHern_2003,Bi2011,Peters_2016} or gelation~\cite{Cates_2004,Zhang2019}, are characterized by strong stress heterogeneities, often referred to as force chains. They are rigid in that they can sustain external shear, yet they are often fragile~\cite{Cates_1998,Bi2011}. Analogous to classical elasticity theory~\cite{Landau_elasticity}, it is plausible to ask whether a long wavelength field theoretic description exists for the mechanical response of such athermal solids and if so, what are its characteristics and universal features, and what would be the {\it appropriate variables} that can account for the underlying kinetic constraints in the emergent field theory? Any attempt to construct such a field theory must answer: (a) how to obtain the stress field, and (b) how to incorporate microscopic information about the structural disorder, accounting for the mechanical constraints into a continuum formulation. This second problem, in particular, has a close resemblance with kinetically constrained models such as hard-core dimer models on lattices where the hard-core constraint of each site being part of one and only one dimer naturally allows for an emergent gauge theory description at low energy and long wavelengths~\cite{PhysRevLett.91.167004,diep2013frustrated,castelnovo2008magnetic}. 

In this Letter, we develop a theory of stress transmission in disordered granular solids, both with and without friction, where the local constraints of mechanical equilibrium are paramount, {\it i.e.} every grain satisfies the constraints of force and torque balance. These local constraints imply that the grain-level stress tensor $\hat\sigma_g$, is symmetric~\cite{Bi2015,DeGiuli2012} and satisfies 
\begin{equation}
( \nabla \cdot \hat{\sigma})_g = \sum_{c \in g} {\bf f}_{g,c} ={\bf f}_{g}^{\rm ext}~.
\label{eq:grain_level_mech_eq}
\end{equation}
Here, $\nabla$ is a discrete divergence operator defined over the underlying contact network, as detailed in the SI~\cite{supplemental_material}, ${\bf f}_{g,c}$ are the contact forces acting on grain $g$ at contact $c$ and ${\bf f}_{g}^{\rm ext}$ is the total external force acting on the grain. Upon coarse-graining~\cite{Ball_Blumenfeld}, Eq.~\eqref{eq:grain_level_mech_eq} gives rise to the continuum condition of mechanical equilibrium: 
\begin{align}
\partial_i \sigma_{ij} ({\bf r})= f_j({\bf r})
\label{eq_mechcont}
\end{align}
where $\sigma_{ij}({\bf r}) $ and $f_j({\bf r})$ are the stress and external force density at the point ${\bf r}$, respectively. 

\addcontentsline{toc}{section}{Stress equation and Tensorial Electromagnetism}
\paragraph*{Stress equation and Tensorial Electromagnetism:} Eq.~\eqref{eq_mechcont}, along with the symmetry of the stress tensor encapsulates the local constraints of mechanical equilibrium. Further, Eq.~\eqref{eq_mechcont} can be casted as the exact analog of Gauss's law:
\begin{eqnarray}
\partial_i E_{ij} &=& \rho_j ,
\label{eq:Gauss}
\end{eqnarray}
in a generalised electromagnetism of $U(1)$ symmetric rank-2 tensor electric fields, $E_{ij}=E_{ji}$, and vector charges, $\rho_i$ with $i=1,\cdots, d$ in $d$ spatial dimensions, the so-called vector charge theory (VCT) of electromagnetism. The resultant generalised Maxwell equations automatically conserve the total charge ($\int d{\bf r} {\boldsymbol {\rho}} =0$) and the dipole moment ($\int d{\bf r} ({\bf r} \times {\boldsymbol { \rho} })=0$)~\cite{Pretko2017a,Pretko2018a}. 

The correspondence between Eq.~\eqref{eq_mechcont} and Eq.~\eqref{eq:Gauss}, along with the conserved quantities makes VCT a natural starting point for deriving the correct continuum theory of the mechanical response of granular solids by formally mapping $E_{ij}\stackrel [] {?} {\leftrightarrow} \sigma_{ij}$ and vector charges to unbalanced forces i.e. ${\boldsymbol {\rho}} \stackrel [] {?}{\leftrightarrow} {\boldsymbol {f}}$. This approach is similar to the problem of frustrated magnets and/or dimer models where due to non-trivial local energetic/kinetic constraints, the individual spins/dimers cease to be the right degrees of freedom and hence fail to describe the low energy theory, which in turn is often described by emergent gauge fields that naturally capture the constraints~\cite{diep2013frustrated,PhysRevLett.91.167004,smbhatt,castelnovo2008magnetic}. Similarly, the displacement of the individual grains from the reference crystalline positions-- the mainstay of the theory of elasticity of crystalline solids~\cite{Ashcroft-Mermin}-- cease to be the right variables in the absence of broken translation symmetry.  However, the long-range stress correlations generated by Newton's laws of force and torque balance are described correctly by the emergent electromagnetism.

%%%%%%%%%%%%%%%%%%%%%%%%%%%%%%%%%%%%%%%%%%%%%%%%
%%%%%%%%%%%%%%%%%%%%%%%%%%%%%%%%%%%%%%%%%%%%%%%%
\begin{figure}[t!]
\hspace{-0.25cm}
\centering
{\includegraphics[width=0.45\textwidth]{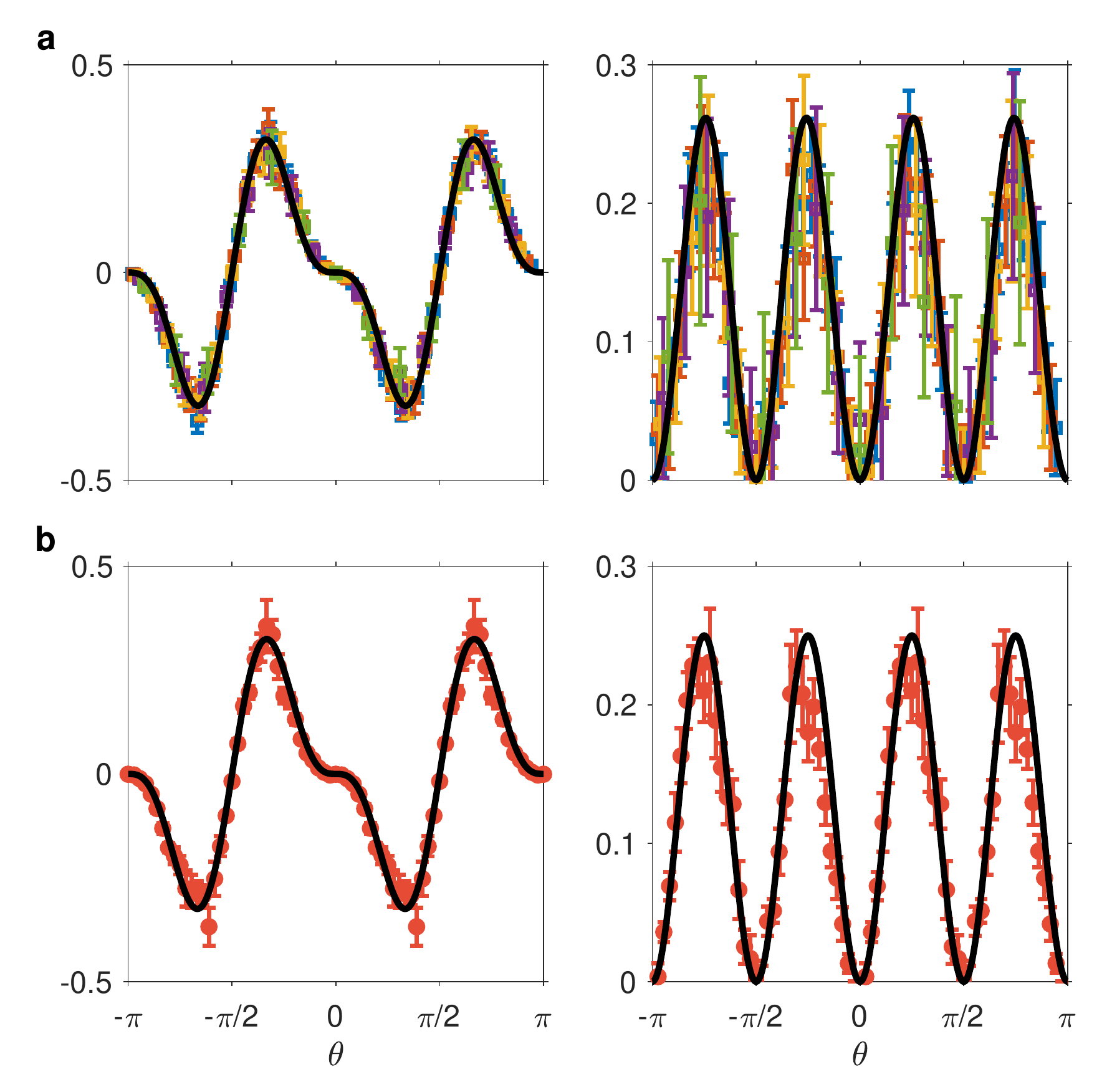}}
\vspace{-0.5cm}
\caption{Comparisons in Fourier space between the theoretical predictions (solid black line) and the disorder-averaged angular dependent stress-stress correlations $C_{xxxy}(\theta)$ and $C_{xxyy}(\theta)$ in the numerical: {\bfseries a}, and the experimental results (red symbols): {\bfseries b}, for isotropically jammed systems. The range of pressure for the numerical data is $P \in [0.016,0.017]$ and the results are displayed for five different system sizes $N = 512, 1024, 2048, 4096, 8192$. The experimental data is from frictional
packings with a range of pressure $P \in [1.5 \times 10^{-4}, 2.9 \times 10^{-4}]$. All correlation functions are normalized by the peak value of $C_{xxxx}(\theta)$. Here, $C_{ijkl} \equiv \langle \sigma_{ij}\sigma_{kl} \rangle$.}
\label{fig:isotropic_angular}
\end{figure}
%%%%%%%%%%%%%%%%%%%%%%%%%%%%%%%%%%%%%%%%%%%%%%%%
%%%%%%%%%%%%%%%%%%%%%%%%%%%%%%%%%%%%%%%%%%%%%%%%

As an immediate consequence of this mapping, we note that the two conservation laws lead to sub-dimensional propagation of charges-- a feature of recently discussed fractonic phases of matter~\cite{Pretko2017a} as well as topological defects in elastic solids~\cite{Pretko2018}. In the present context, it also provides a natural explanation for the visually striking ``force chains'' (see Fig. \ref{fig:sheared}) observed in photoelastic images of granular solids~\cite{Trush_isotropic}, as our analysis will demonstrate.

%%%%%%%%%%%%%%%%%%%%%%%%%%%%%%%%%%%%%%%%%%%%%%%%
%%%%%%%%%%%%%%%%%%%%%%%%%%%%%%%%%%%%%%%%%%%%%%%%
\begin{figure*}[t!]
%\hspace{-0.25cm}
\centering
{\includegraphics[width=0.8\textwidth]{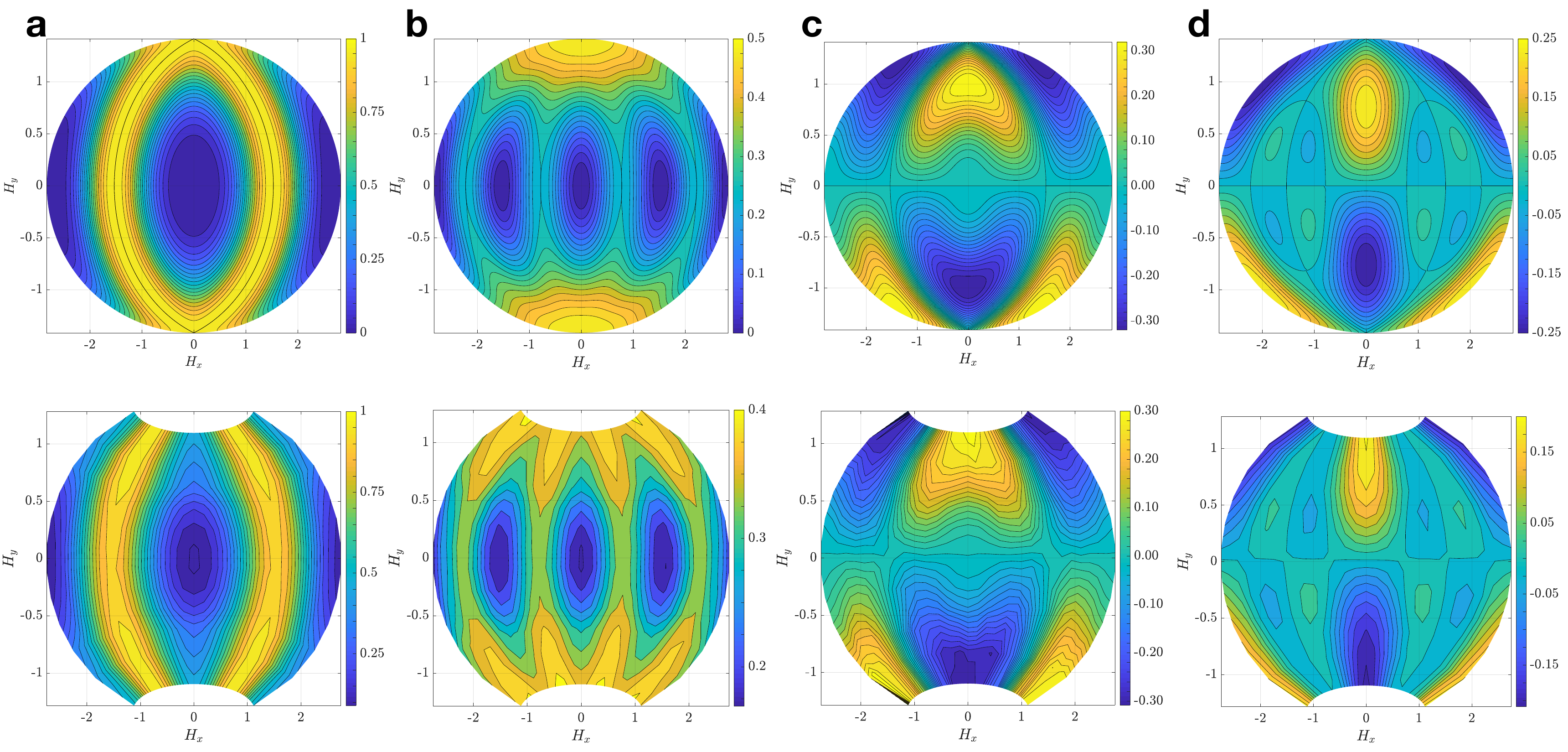}}
%\vspace{-0.75cm}
\caption{Comparisons in Fourier space between theoretical predictions (top) and numerical results (bottom) from jammed packings of frictionless spheres in three dimensions. The figures display the radially averaged correlation functions {\bf a}: $C_{xxxx}(\theta,\phi)$, {\bf b}: $C_{xyxy}(\theta,\phi)$, {\bf c}: $C_{xxxz}(\theta,\phi)$ and {\bf d}: $C_{xyyz}(\theta,\phi)$ respectively. The coordinates ($H_x, H_y$) represent a Hammer projection of the $(\theta,\phi)$ shell onto the plane. The results are {presented} for system size $N=27000$, {and have been} averaged over $350$ configurations. The range of packing fractions for these configurations is $\phi \in [0.686,0.689]$ and the range of pressure per grain is $P \in [0.0136,0.0147]$. Results for the $\hat{\Lambda}$ tensor have not been presented due to the small effective system size: $30 \times 30 \times 30$. The blank regions at the poles $\phi$ at $\theta=0$ and $\theta=\pi$ in the numerical results is due to the difficulty in sampling these points.} 
\label{fig:3D}
\end{figure*}
%%%%%%%%%%%%%%%%%%%%%%%%%%%%%%%%%%%%%%%%%%%%%%%%
%%%%%%%%%%%%%%%%%%%%%%%%%%%%%%%%%%%%%%%%%%%%%%%%

It is well known~\cite{Bouchaud-ideas} that Eq.~\eqref{eq_mechcont} does not provide enough equations to solve for the field $\sigma_{ij}$, since there are only $d$ equations for the $d(d+1)/2$ components of a symmetric tensor in $d$ dimensions~\cite{Pretko2017a}. These missing equations are provided within VCT, by invoking the complete set of Maxwell's equations required to uniquely specify $E_{ij}$. In particular, the generalised Faraday's law: $ \frac{\partial B_{ij}}{\partial t} = - \epsilon_{iab} \epsilon_{jcd} \partial_a \partial_c E_{bd}$, where $B_{ij}=B_{ji}$ is the tensor magnetic field of VCT, leads to the generalised irrotational condition $\epsilon_{iab} \epsilon_{jcd} \partial_a \partial_c E_{bd}=0$ in the electrostatic limit. This condition provides the missing equations, and leads to the potential formulation: $E_{ij} = \frac{1}{2} (\partial_i \phi_j + \partial_j \phi_i )$, where $\phi_i$ is the electrostatic potential which can be used to obtain $E_{ij}$ for {\it any} charge configuration~\cite{Pretko2017a}.

\paragraph*{Granular solid as a generalised dielectric medium:} \addcontentsline{toc}{section}{Granular solid as a generalised dielectric medium} The above gauge theory formulation containing all the basic ingredients, requires an extension-- akin to that of dielectric media-- to capture the complexity of the granular mechanics. This is easily seen by noting the twin crucial characteristics of granular media: (1) it is {\it only} defined under external pressure (as a packing of grains with purely repulsive interactions will fall apart in the absence of boundary forces), and (2) it can support internal stresses. This translates, within VCT, to an assembly being subject to well defined boundary charges developing internal {\it charge dipoles}, akin to the response of a polarizable medium (dielectric). Alternatively stated, although a granular solid under external compression is free of ``charges'' since every grain satisfies force and torque balance, ``bound charges'' exist as pairs of equal and opposite forces at every contact of the disordered granular network. Hence we re-write Eq.~\eqref{eq:Gauss} as

\begin{eqnarray}
\partial_i E_{ij} &=& \rho_j^{free} + \rho_j^{bound}, 
\label{eq:pol}
\end{eqnarray}  
where $\rho_j^{free}$ arises from any body-force such as gravity and $\rho_j^{bound}$ are the bound charges arising from the force-dipoles and can be accounted for using a tensorial dipole moment $P_{ij}$ such that
\begin{align}
\partial_i P_{ij} =- \rho_j^{bound}
\end{align}
 A complete derivation of these relations, and a detailed discussion of the structure of the theory will be presented in a future paper. The structure of $P_{ij}$ is influenced by various microscopic details of the system such as the features of the underlying contact network and the nature of contact forces which, for example can be purely repulsive or both repulsive and attractive and frictionless or frictional. 
 
To construct a continuum theory, we assert that $P_{ij}$ is related to $E_{ij}$ through a fourth-rank polarizability tensor, $\chi_{ijkl}$, as in linear dielectrics : $P_{ij} = \chi_{ijkl} E_{kl}$. Straightforward generalisation of electrostatics in dielectric medium follows. We define a ``Displacement" tensor, 
 \begin{align}
 D_{ij}= {{(\delta_{ik}\delta_{jl}}} + \chi_{ijkl}) E_{kl}=(\Lambda^{-1})_{ijkl}E_{kl},
 \label{eq_lambda}
 \end{align}
that satisfies
\begin{align}
\partial_i D_{ij} = \rho_j^{free}; ~ ~ ~\epsilon_{iab} \epsilon_{jcd} \partial_a \partial_c (\Lambda D)_{bd} = 0 ~.
\label{eq:statics}
\end{align}
The inverse dielectric tensor, $\Lambda$, satisfies $\Lambda_{ijkl} = \Lambda_{jikl} = \Lambda_{ijlk} = \Lambda_{jilk}$. Since the inherent stresses in a granular solid satisfy the first relation in Eq.~\eqref{eq:statics} as a direct consequence of force balance, we interpret $D_{ij}$ as the Cauchy stress tensor measured from contact forces and contact vectors inside the material, {\it i.e.}\ {$D_{ij} \leftrightarrow \sigma_{ij}$} in Eq.~\eqref{eq:statics}.

Eq.~\eqref{eq:statics}, which is our main theoretical result, can be compared to anisotropic elasticity~\cite{Otto2003}:
\begin{eqnarray}
\partial_i \sigma_{ij} &=&0~; ~ ~ \epsilon_{iab} \epsilon_{jcd} \partial_a \partial_c U_{bd} = 0, \nonumber \\
\sigma_{ij} &=& ( \Xi)^{-1}_{ijkl}U_{kl}.
\label{eq:aniso}
\end{eqnarray}
where $U_{ij}$ is the macroscopic strain tensor. In Eq.~\eqref{eq:aniso},  $\hat{\Xi}(\leftrightarrow \Lambda ~\rm{appearing ~ in ~Eq. ~\eqref{eq:statics}})$  is the inverse of the elastic modulus tensor. Identifying {$E_{ij} \leftrightarrow U_{ij}$, and $D_{ij} \leftrightarrow \sigma_{ij}$}, demonstrates that an elasticity theory capturing the stress responses in granular solids~\cite{Otto2003,Lemaitre2018,Geng2001} emerges from VCT. A gauge potential, ${\bm \phi}$ replaces the displacement field in elasticity theory. We would like to reiterate that unlike the displacement field in elasticity, the gauge potential, $\boldsymbol{\phi}$, is not a physical observable. Therefore, VCT provides the ``missing'' compatibility equations that allow us to solve the granular stress response problem without invoking a displacement field. Thus, although $\hat{D}$ and $\hat{E}$ have a correspondence with $\hat{\sigma}$ and $\hat{U}$, the elasticity emerges from local constraints and not from broken symmetry. This {\it stress-only} description does not refer to a stress-free state or displacement fields. Moreover, the effective elastic modulus, $\hat{\Lambda}$, is not constrained by symmetries imposed by a free-energy and will depend on protocols.

We note that the missing equation in 2D had been obtained for the {\it particular} contact geometry of hard-particle frictional jammed states~\cite{Ball_Blumenfeld} by introducing a geometry-related symmetric tensor. The relation of this network-specific description to VCT needs to be explored further. However, the potential formulation~\cite{Ball_Blumenfeld} is identical to the dual representation of VCT in 2D~\cite{Xu2013,supplemental_material}.

 %%%%%%%%%%%%%%%%%%%%%%%%%%%%%%%%%%%%%%%%%%%%%%%%
%%%%%%%%%%%%%%%%%%%%%%%%%%%%%%%%%%%%%%%%%%%%%%%%
\begin{figure*}[t!]
%\hspace{-0.25cm}
\centering
{\includegraphics[width=0.8\textwidth]{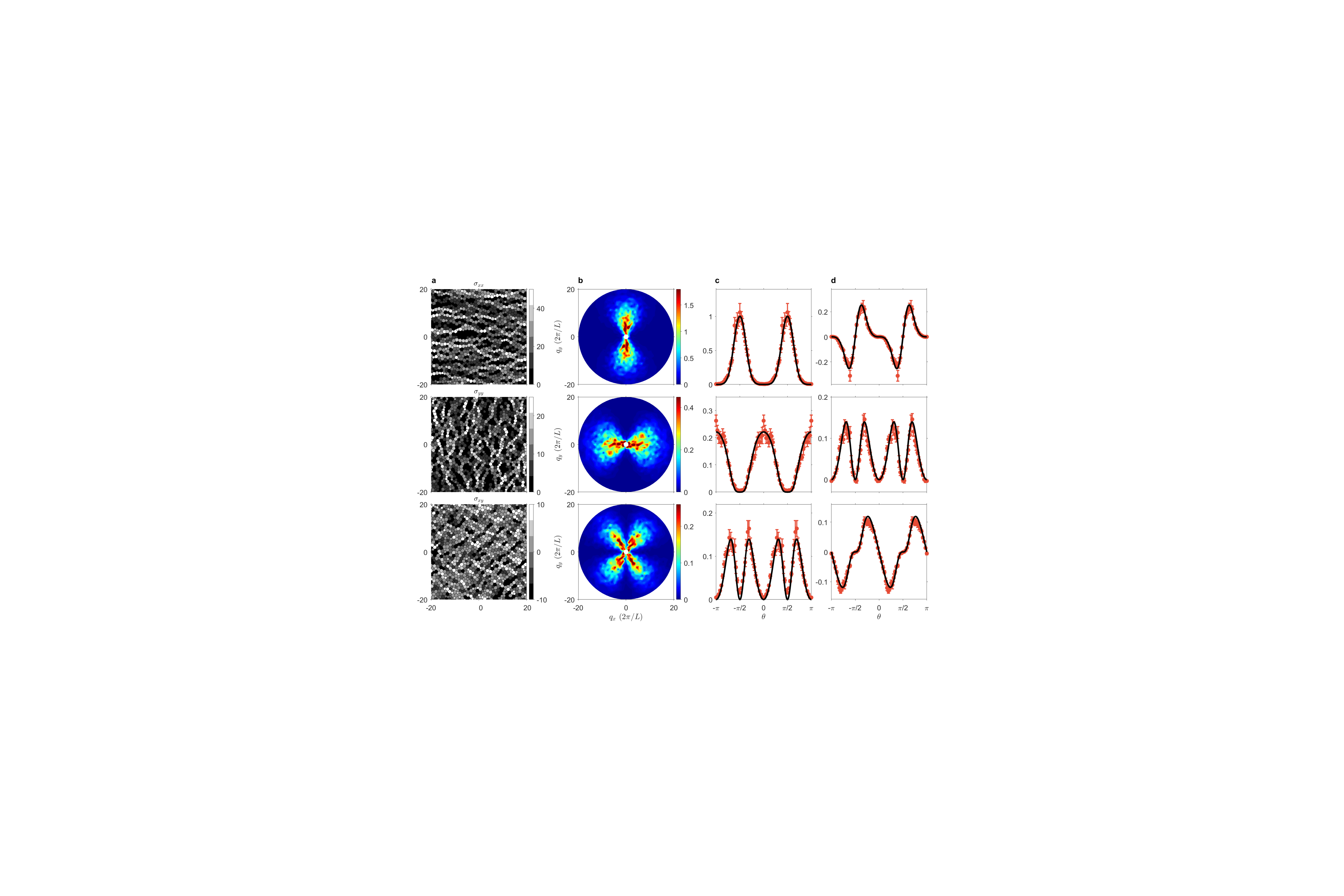}}
%\vspace{-0.75cm}
\caption{{Comparisons in Fourier space between the experimental results (red symbols) in {\it sheared} frictional packings and the theoretical predictions (black line)}. {\bfseries a}: Photoelastic images produced from a sheared packing. {\bfseries b}: Contour plots of the stress-stress correlation functions: $C_{xxxx}(q,\theta)$ ({\bfseries Top}), $C_{yyyy}(q,\theta)$ ({\bfseries Middle}) and $C_{xyxy}(q,\theta)$ ({\bfseries Bottom}). {\bfseries c}: Corresponding angular plots. {\bfseries d}: Angular plots of $C_{xxxy}(\theta)$ ({\bfseries Top}), $C_{xxyy}(\theta)$ ({\bfseries Middle}) and $C_{xyyy}(\theta)$ ({\bfseries Bottom}). All correlations are normalized by the peak value of $C_{xxxx}(\theta)$. The $\hat{\Lambda}$ tensor obtained from the fit is diagonal with $\lambda_{11}=1$, $\lambda_{22} = 4.5$, and $\lambda_{33}=1.46$. The ratio of the boundary stress components defining the shear is $\Sigma_{xx}/\Sigma_{yy} = 1.94$, which satisfies the positivity bound~\cite{Henkes2009,Lois2009} $\frac{\lambda_{22}}{\lambda_{11}} \ge (\frac{\Sigma_{xx}}{\Sigma_{yy}})^2$.}
\label{fig:sheared}
\end{figure*}
%%%%%%%%%%%%%%%%%%%%%%%%%%%%%%%%%%%%%%%%%%%%%%%%
%%%%%%%%%%%%%%%%%%%%%%%%%%%%%%%%%%%%%%%%%%%%%%%%

\paragraph*{Experiments and numerical results:} \addcontentsline{toc}{section}{Experiments and numerical results} 
 We have compared the predictions for stress-stress correlations obtained from Eq.~\eqref{eq:statics} to experimental and numerical data, and extracted $\hat{\Lambda}$ for frictionless and frictional granular solids prepared under different protocols. A hallmark of the VCT in free space (Eq.~\eqref{eq:Gauss}), both in 2D and 3D, is the appearance of pinch point singularities in the Fourier transforms of $E_{ij}$ correlators~\cite{Prem2018}:

\begin{widetext}
\begin{align}
C^{free}_{ijkl} ({\bf q}) & \equiv \left\langle E_{ij}({\bf q}) E_{kl}(-{\bf q})\right\rangle \ \propto \frac{\left(\delta_{ik} \delta_{jl} +\delta_{il} \delta_{jk}\right)}{2} +\frac{q_i q_j q_k q_l}{q^4} - \frac{1}{2}\left(\frac{\delta_{ik} q_j q_l}{q^2}+\frac{\delta_{jk} q_i q_l}{q^2}+\frac{\delta_{il} q_j q_k}{q^2}+\frac{\delta_{jl} q_i q_k}{q^2}\right) ~.
\label{eq:Ecorr}
\end{align}
\end{widetext}
Eq.~\eqref{eq:Ecorr} is obtained by imposing the Gauss's law constraint, $\partial_i E_{ij} = 0$, on Eq.~\eqref{eq:Gauss}, and assuming that all states are equiprobable~\cite{Prem2018}, i.e.\ the Edwards measure~\cite{Bi2015}. Earlier granular field theories~\cite{Henkes2009, Lois2009, deGiuli_PRL} based on this measure used a dual formulation of VCT~\cite{Xu2013,Ball_Blumenfeld} where the emergence of elasticity is not evident. Since $C^{free}_{ijkl} ({\bf q})$ is independent of $|{\bf q}|$, it is straightforward to show that the correlations in real-space decay as $1/r^d$. A more stringent test of the theory, therefore is the pinch-point structure of the correlation functions.

For granular solids, we computed
the correlators $C_{ijkl} ({\bf q})=\left \langle D_{ij}({\bf q}) D_{kl}(-{\bf q})\right\rangle$ using Eq.~\eqref{eq:statics}, and tested the predictions in ensembles of 2D and 3D isotropically compressed soft particles (numerically), and in ensembles of 2D packings of frictional grains (experimentally). {Pinch point singularities are clearly exhibited in both 2D (Figs. \ref{fig:isotropic_angular} and \ref{fig:sheared}) and 3D (Fig. \ref{fig:3D}).}
We have determined $\hat{\Lambda}$ through detailed comparisons between theory and measurements of $C_{ijkl}$ (Figs. \ref{fig:isotropic_angular} - \ref{fig:sheared}). Fig. \ref{fig:isotropic_angular} demonstrates that for packings created under isotropic compression, $\hat{\Lambda} = \lambda {\mathbb{1}}$, with ${\mathbb{1}}$ being the identity tensor. Additional  tests of the theory are presented in the SI~\cite{supplemental_material}.

To illustrate the sensitivity of the $\hat{\Lambda}$ tensor to protocol (stress ensemble) we generated sheared packings of the same {grains used} in the isotropic compression. Under the experimental conditions of pure shear, with principle stress along $x$ and $y$, a diagonal form with different values of $\lambda_{ii}$ provides an excellent description of the experimental observations (Fig. \ref{fig:sheared}). We find that $\lambda_{ii}$, satisfy a set of bounds imposed by the constraint of positivity of normal forces in granular media~\cite{Henkes2009,Lois2009}. 

A consequence of the pinch point singularities is that in real-space, $\langle D_{ij} ({\bf q}) D_{ij} (-{\bf q}) \rangle$ is negative in transverse directions and positive along longitudinal directions, as shown in the SI~\cite{supplemental_material}. It is this property that is strikingly demonstrated in photoelastic images of force chains in Fig. \ref{fig:sheared} and Figs. 5, 6, and 7 in the SI~\cite{supplemental_material}.

\paragraph*{Summary and Outlook:}\addcontentsline{toc}{section}{Summary and Outlook}
 To summarize, we have demonstrated that the elasticity of athermal amorphous solids is described by an exact analog of the electrostatics of a fractonic $U(1)$ gauge theory~\cite{Pretko2017a} in polarizable media. Although our analysis focused on granular solids, it marks a paradigm shift in our understanding of ``amorphous elasticity'' for a much broader class of solids such as jammed suspensions~\cite{Peters_2016}, disordered crystals~\cite{Pappu2019} and colloidal gels~\cite{Zhang2019} with no broken symmetry but strictly enforced {\it local} constraints of force {and} torque balance. A key to the above approach is the lack of a {\it reference} configuration rendering $\boldsymbol{\phi}$ to be un-observable. In a crystal, this is no longer true. Further, the crystal can survive in a zero-stress condition. Based on these observations, our preliminary considerations suggest that the crystalline elasticity due to broken symmetry modifies the nature of the media, possibly involving a Higgs mechanism \textit{i.e. a crystal is a momentum condensate}~\cite{Phil_Anderson}. In such a scenario, the plasmon excitations of the tensor electromagnetism would appear as optical phonons. However, the emergence of gapless acoustic phonons currently remains unclear.

%\txtbl{This general framework should lead to crystalline elasticity upon incorporation of broken symmetry, possibly through a Higgs mechanism transforming the gauge potentials to physical fields.}

The theory can be extended to stress-correlations in thermal amorphous solids such as low-temperature glass formers~\cite{Gelin2016,Maier2017,PhysRevE.91.032301}. As in frustrated magnets~\cite{diep2013frustrated}, thermal fluctuations lead to a length-scale characterizing the distance between particles at which force and torque balance are violated, and wash out the pinch-point singularity~\cite{supplemental_material}. This low-temperature extension does not include the physics of the glass transition, which addresses the onset of rigidity in supercooled liquids. The theory we have presented assumes that all disordered networks that satisfy the constraints of mechanical equilibrium are equiprobable. If and why the glass transition generates this ensemble is a an important question that we have not addressed.

A fully dynamical theory of amorphous materials can be constructed by extending the ``electrostatics'' to ``electrodynamics'' through the identification of the analog of a magnetic field, and including unbalanced forces as charged excitations~\cite{Pretko2018a}.

\paragraph*{Acknowledgements:} \addcontentsline{toc}{section}{Acknowledgements}
The authors acknowledge Michael D'Eon, Albion Lawrence, Debarghya Banerjee, Chandan Dasgupta, Kedar Damle, Roderich Moessner, Nandagopal M, Prasad Perlekar, Samriddhi Sankar Ray and Vijay Shenoy for fruitful discussions. JN and BC acknowledge support from NSF-CBET-1916877 and BSF-2016188. BC was supported by a Simons Fellowship in Theoretical Physics. SB acknowledges financial support of Max Planck partner group on strongly correlated systems at ICTS; SERB-DST (Govt. of India) Early Career Research grant (No. ECR/2017/000504) and the Department of Atomic Energy (DAE), Govt. of India, under project no.12-R\&D- TFR-5.10-1100. YW and JZ acknowledge the support from NSFC under (No. 11774221 and No. 11974238). BC, JN and SB would like to thank the ICTS program: Entropy, Information and Order in Soft Matter (August-November 2018) during which a part of the work was done. The work of JN and KR was funded in part by intramural funds at TIFR Hyderabad from the DAE.

%\bibliography{granular_stress.bib} 

\begin{thebibliography}{10}
\expandafter\ifx\csname url\endcsname\relax
 \def\url#1{\texttt{#1}}\fi
\expandafter\ifx\csname urlprefix\endcsname\relax\def\urlprefix{URL }\fi
\providecommand{\bibinfo}[2]{#2}
\providecommand{\eprint}[2][]{\url{#2}}

\bibitem{OHern_2003}
\bibinfo{author}{O'Hern, C.~S.}, \bibinfo{author}{Silbert, L.~E.},
 \bibinfo{author}{Liu, A.~J.} \& \bibinfo{author}{Nagel, S.~R.}
\newblock \bibinfo{title}{Jamming at zero temperature and zero applied stress:
 The epitome of disorder}.
\newblock \emph{\bibinfo{journal}{Phys. Rev. E}} \textbf{\bibinfo{volume}{68}},
 \bibinfo{pages}{011306} (\bibinfo{year}{2003}).

\bibitem{Bi2011}
\bibinfo{author}{Bi, D.}, \bibinfo{author}{Zhang, J.},
 \bibinfo{author}{Chakraborty, B.} \& \bibinfo{author}{Behringer, R.~P.}
\newblock \bibinfo{title}{{Jamming by shear}}.
\newblock \emph{\bibinfo{journal}{Nature}} \textbf{\bibinfo{volume}{480}},
 \bibinfo{pages}{355} (\bibinfo{year}{2011}).
 
 \bibitem{Cates_1998}
\bibinfo{author}{Cates, M.~E.}, \bibinfo{author}{Wittmer, J.~P.},
 \bibinfo{author}{Bouchaud, J.-P.} \& \bibinfo{author}{Claudin, P.}
\newblock \bibinfo{title}{{Jamming, force chains, and fragile matter}}.
\newblock \emph{\bibinfo{journal}{Phys. Rev. Lett.}}
 \textbf{\bibinfo{volume}{81}}, \bibinfo{pages}{1841--1844}
 (\bibinfo{year}{1998}).

\bibitem{Peters_2016}
\bibinfo{author}{Peters, I.~R.}, \bibinfo{author}{Majumdar, S.} \&
 \bibinfo{author}{Jaeger, H.~M.}
\newblock \bibinfo{title}{{Direct observation of dynamic shear jamming in dense
 suspensions}}.
\newblock \emph{\bibinfo{journal}{Nature}} \textbf{\bibinfo{volume}{532}},
 \bibinfo{pages}{214} (\bibinfo{year}{2016}).

 \bibitem{Cates_2004}
\bibinfo{author}{Cates, M.~E.}, \bibinfo{author}{Fuchs, M.},
 \bibinfo{author}{Kroy, K.}, \bibinfo{author}{Poon, W. C.~K.} \&
 \bibinfo{author}{Puertas, A.~M.}
\newblock \bibinfo{title}{{Theory and simulation of gelation, arrest and
 yielding in attracting colloids}}.
\newblock \emph{\bibinfo{journal}{Journal of Physics: Condensed Matter}}
 \textbf{\bibinfo{volume}{16}}, \bibinfo{pages}{S4861}
 (\bibinfo{year}{2004}).


\bibitem{Zhang2019}
\bibinfo{author}{Zhang, S.} \emph{et~al.}
\newblock \bibinfo{title}{{Correlated Rigidity Percolation and Colloidal
 Gels}}.
\newblock \emph{\bibinfo{journal}{Physical Review Letters}}
 \textbf{\bibinfo{volume}{123}}, \bibinfo{pages}{33--35}
 (\bibinfo{year}{2019}).

\bibitem{Landau_elasticity}
\bibinfo{author}{Landau, L.~D.}, \bibinfo{author}{Pitaevskii, L.~P.},
 \bibinfo{author}{Kosevich, A.~M.} \& \bibinfo{author}{Lifshitz, E.}
\newblock \emph{\bibinfo{title}{Theory of Elasticity}}
 (\bibinfo{publisher}{Elsevier}, \bibinfo{year}{2012}).
 
 
 \bibitem{diep2013frustrated}
\bibinfo{author}{Diep, H.} \emph{et~al.}
\newblock \emph{\bibinfo{title}{Frustrated spin systems}}
 (\bibinfo{publisher}{World Scientific}, \bibinfo{year}{2013}).

\bibitem{PhysRevLett.91.167004}
\bibinfo{author}{Huse, D.~A.}, \bibinfo{author}{Krauth, W.},
 \bibinfo{author}{Moessner, R.} \& \bibinfo{author}{Sondhi, S.~L.}
\newblock \bibinfo{title}{Coulomb and liquid dimer models in three dimensions}.
\newblock \emph{\bibinfo{journal}{Phys. Rev. Lett.}}
 \textbf{\bibinfo{volume}{91}}, \bibinfo{pages}{167004}
 (\bibinfo{year}{2003}).


\bibitem{castelnovo2008magnetic}
\bibinfo{author}{Castelnovo, C.}, \bibinfo{author}{Moessner, R.} \&
 \bibinfo{author}{Sondhi, S.~L.}
\newblock \bibinfo{title}{Magnetic monopoles in spin ice}.
\newblock \emph{\bibinfo{journal}{Nature}} \textbf{\bibinfo{volume}{451}},
 \bibinfo{pages}{42} (\bibinfo{year}{2008}).
 

\bibitem{Bi2015}
\bibinfo{author}{Bi, D.}, \bibinfo{author}{Henkes, S.},
 \bibinfo{author}{Daniels, K.~E.} \& \bibinfo{author}{Chakraborty, B.}
\newblock \bibinfo{title}{{The Statistical Physics of Athermal Materials}}.
\newblock \emph{\bibinfo{journal}{Annual Review of Condensed Matter Physics}}
 \textbf{\bibinfo{volume}{6}}, \bibinfo{pages}{63--83} (\bibinfo{year}{2015}).

\bibitem{DeGiuli2012}
\bibinfo{author}{DeGiuli, E.}
\newblock \bibinfo{title}{{Continuum limits of granular systems}}
(\bibinfo{publisher}{Thesis, U. British Columbia}, (\bibinfo{year}{2012}).

  
 \bibitem{Ball_Blumenfeld} 
\bibinfo{author}{Ball, Robin C.} \& \bibinfo{author}{Blumenfeld, Raphael,} \newblock \bibinfo{title}{{Stress Field in Granular Systems: Loop Forces and Potential Formulation}}.
\newblock \emph{\bibinfo{journal}{Physical Review Letters}}
\textbf{\bibinfo{volume}{88}},\bibinfo{pages}{115505} (\bibinfo{year}{2002}).

\bibitem{Pretko2017a}
\bibinfo{author}{Pretko, M.}
\newblock \bibinfo{title}{{Generalized electromagnetism of subdimensional
 particles: A spin liquid story}}.
\newblock \emph{\bibinfo{journal}{Phys. Rev. B}}
 \textbf{\bibinfo{volume}{96}}, \bibinfo{pages}{1--26} (\bibinfo{year}{2017}).

\bibitem{Pretko2018a}
\bibinfo{author}{Pretko, M.}
\newblock \bibinfo{title}{{The fracton gauge principle}}.
\newblock \emph{\bibinfo{journal}{Phys. Rev. B}}
 \textbf{\bibinfo{volume}{98}}, \bibinfo{pages}{115134} (\bibinfo{year}{2018}).
 
 \bibitem{smbhatt}
\bibinfo{author}{Domb, C.}, \bibinfo{author}{Green, M.S.} \&
 \bibinfo{author}{Lebowitz, J.L.}
\newblock \emph{\bibinfo{title}{Phase Transitions and Critical Phenomena Vol.
 13 edited by by C. Domb and J. Lebowitz}} (\bibinfo{publisher}{Academic
 Press, London}, \bibinfo{year}{1989}). 

\bibitem{Ashcroft-Mermin}
\bibinfo{author}{Ashcroft, N.~W.} \& \bibinfo{author}{Mermin, N.~D.}
\newblock \emph{\bibinfo{title}{Solid State Physics}}
 (\bibinfo{publisher}{Saunders College Publishing}, \bibinfo{year}{1976}).

\bibitem{Pretko2018}
\bibinfo{author}{Pretko, M.} \& \bibinfo{author}{Radzihovsky, L.}
\newblock \bibinfo{title}{{Fracton-Elasticity Duality}}.
\newblock \emph{\bibinfo{journal}{Phys. Rev. Lett. }}
 \textbf{\bibinfo{volume}{120}}, \bibinfo{pages}{195301} (\bibinfo{year}{2018}).

\bibitem{Trush_isotropic}
\bibinfo{author}{Majmudar, T.~S.} \& \bibinfo{author}{Behringer, R.~P.}
\newblock \bibinfo{title}{Contact force measurements and stress-induced
 anisotropy in granular materials}.
\newblock \emph{\bibinfo{journal}{Nature}} \textbf{\bibinfo{volume}{435}},
 \bibinfo{pages}{1079--1082} (\bibinfo{year}{2005}).
 

\bibitem{Bouchaud-ideas}
\bibinfo{author}{Bouchaud, J.-P.}
\newblock \bibinfo{title}{Granular media: some ideas from statistical physics}.
\newblock In \bibinfo{editor}{Bouchaud, J.}, \bibinfo{editor}{Barrat, J.~L.},
 \bibinfo{editor}{Feigelman, M.}, \bibinfo{editor}{Kurchan, J.} \&
 \bibinfo{editor}{Dalibard, J.} (eds.) \emph{\bibinfo{booktitle}{Slow
 Relaxations and Nonequilibrium Dynamics in Condensed Matter}},
 vol.~\bibinfo{volume}{77}, \bibinfo{pages}{185--202} (\bibinfo{publisher}{Les
 Ulis: EDP Sciences}, \bibinfo{year}{2003}).

\bibitem{Otto2003}
\bibinfo{author}{Otto, M.}, \bibinfo{author}{Bouchaud, J.~P.},
 \bibinfo{author}{Claudin, P.} \& \bibinfo{author}{Socolar, J.~E.}
\newblock \bibinfo{title}{{Anisotropy in granular media: Classical elasticity
 and directed-force chain network}}.
\newblock \emph{\bibinfo{journal}{Phys. Rev. E}}
 \textbf{\bibinfo{volume}{67}}, \bibinfo{pages}{24} (\bibinfo{year}{2003}).


\bibitem{Lemaitre2018}
\bibinfo{author}{Lema{\^{i}}tre, A.}
\newblock \bibinfo{title}{{Stress correlations in glasses}}.
\newblock \emph{\bibinfo{journal}{Journal of Chemical Physics}}
 \textbf{\bibinfo{volume}{149}}, \bibinfo{pages}{104107}
 (\bibinfo{year}{2018}).

\bibitem{Geng2001}
\bibinfo{author}{Geng, J.} \emph{et~al.}
\newblock \bibinfo{title}{{Footprints in sand: The response of a granular
 material to local perturbations}}.
\newblock \emph{\bibinfo{journal}{Phys. Rev. Lett. }}
 \textbf{\bibinfo{volume}{87}}, \bibinfo{pages}{035506}
 (\bibinfo{year}{2001}).
 
\bibitem{Xu2013}
\bibinfo{author}{Xu, C.}
\newblock \bibinfo{title}{{Novel Algebraic Boson Liquid phase with soft
 Graviton excitations}} 
 \newblock \emph{\bibinfo{journal}{arXiv: 0602443v4}} (\bibinfo{year}{2006}).
 
\bibitem{Prem2018}
\bibinfo{author}{Prem, A.}, \bibinfo{author}{Vijay, S.}, \bibinfo{author}{Chou,
 Y.~Z.}, \bibinfo{author}{Pretko, M.} \& \bibinfo{author}{Nandkishore, R.~M.}
\newblock \bibinfo{title}{{Pinch point singularities of tensor spin liquids}}.
\newblock \emph{\bibinfo{journal}{Phys.Rev. B}}
 \textbf{\bibinfo{volume}{98}}, \bibinfo{pages}{1} (\bibinfo{year}{2018}).

 
 \bibitem{Henkes2009}
\bibinfo{author}{Henkes, S.} \& \bibinfo{author}{Chakraborty, B.}
\newblock \bibinfo{title}{{Statistical mechanics framework for static granular
 matter}}.
\newblock \emph{\bibinfo{journal}{Phys. Rev. E }} \textbf{\bibinfo{volume}{79}},
\bibinfo{pages}{061301} (\bibinfo{year}{2009}).

\bibitem{deGiuli_PRL}
\bibinfo{author}{DeGiuli, E.}
\newblock \bibinfo{title}{{Field Theory for Amorphous Solids}}.
\newblock \emph{\bibinfo{journal}{Phys. Rev. Lett.}}
 \textbf{\bibinfo{volume}{121}}, \bibinfo{pages}{118001}
 (\bibinfo{year}{2018}). 
 
 
\bibitem{Lois2009}
\bibinfo{author}{Lois, G.} \emph{et~al.}
\newblock \bibinfo{title}{{Stress correlations in granular materials: An
 entropic formulation}}.
\newblock \emph{\bibinfo{journal}{Phys. Rev. E }} \textbf{\bibinfo{volume}{80}}, \bibinfo{pages}{060303(R)}
 (\bibinfo{year}{2009}).


\bibitem{Pappu2019}
\bibinfo{author}{Acharya, P.}, \bibinfo{author}{Sengupta, S.},
 \bibinfo{author}{Chakraborty, B.} \& \bibinfo{author}{Ramola, K.}
\newblock \bibinfo{title}{Athermal fluctuations in disordered crystals}
 \newblock \emph{\bibinfo{journal}{arXiv: 1910.06352}} (\bibinfo{year}{2019}).
 

 \bibitem{Phil_Anderson}
\bibinfo{author}{Anderson, P.}, 
\newblock \emph{\bibinfo{title}{Basic Notions Of Condensed Matter Physics}} (\bibinfo{publisher}{CRC Press}, \bibinfo{year}{1994})


\bibitem{Gelin2016}
\bibinfo{author}{Gelin, S.}, \bibinfo{author}{Tanaka, H.} \&
 \bibinfo{author}{Lema{\^{i}}tre, A.}
\newblock \bibinfo{title}{{Anomalous phonon scattering and elastic correlations
 in amorphous solids}}.
\newblock \emph{\bibinfo{journal}{Nature Materials}}
 \textbf{\bibinfo{volume}{15}}, \bibinfo{pages}{1177}
 (\bibinfo{year}{2016}).

\bibitem{Maier2017}
\bibinfo{author}{Maier, M.}, \bibinfo{author}{Zippelius, A.} \&
 \bibinfo{author}{Fuchs, M.}
\newblock \bibinfo{title}{{Emergence of Long-Ranged Stress Correlations at the
 Liquid to Glass Transition}} \newblock \emph{\bibinfo{journal}{Phys. Rev. Lett.}} \textbf{\bibinfo{volume}{119}},
 \bibinfo{pages}{265701} (\bibinfo{year}{2017}).


\bibitem{PhysRevE.91.032301}
\bibinfo{author}{Wu, B.}, \bibinfo{author}{Iwashita, T.} \&
 \bibinfo{author}{Egami, T.}
\newblock \bibinfo{title}{Anisotropic stress correlations in two-dimensional
 liquids}.
\newblock \emph{\bibinfo{journal}{Phys. Rev. E}} \textbf{\bibinfo{volume}{91}},
 \bibinfo{pages}{032301} (\bibinfo{year}{2015}).

  \bibitem{supplemental_material} 
  {See Supplemental Material for details, which includes Refs. [35-44].}

%%Following are references exclusively used in SI

\bibitem{ohern1}
\bibinfo{author}{O'Hern, C.~S.}, \bibinfo{author}{Langer, S.~A.},
 \bibinfo{author}{Liu, A.~J.} \& \bibinfo{author}{Nagel, S.~R.}
\newblock \bibinfo{title}{Random packings of frictionless particles}.
\newblock \emph{\bibinfo{journal}{Phys. Rev. Lett.}}
 \textbf{\bibinfo{volume}{88}}, \bibinfo{pages}{075507}
 (\bibinfo{year}{2002}).

\bibitem{Ramola2017}
\bibinfo{author}{Ramola, K.} \& \bibinfo{author}{Chakraborty, B.}
\newblock \bibinfo{title}{{Stress Response of Granular Systems}}.
\newblock \emph{\bibinfo{journal}{Journal of Statistical Physics}}
 \textbf{\bibinfo{volume}{169}} (\bibinfo{year}{2017}).

\bibitem{FIRE}
\bibinfo{author}{Bitzek, E.}, \bibinfo{author}{Koskinen, P.},
 \bibinfo{author}{G\"ahler, F.}, \bibinfo{author}{Moseler, M.} \&
 \bibinfo{author}{Gumbsch, P.}
\newblock \bibinfo{title}{Structural relaxation made simple}.
\newblock \emph{\bibinfo{journal}{Phys. Rev. Lett.}}
 \textbf{\bibinfo{volume}{97}}, \bibinfo{pages}{170201}
 (\bibinfo{year}{2006}).

\bibitem{FIRE2.0}
\bibinfo{author}{Gu{\'e}nol{\'e}, J.} \emph{et~al.}
\newblock \bibinfo{title}{Assessment and optimization of the fast inertial
 relaxation engine (fire) for energy minimization in atomistic simulations and
 its implementation in lammps} \newblock \emph{\bibinfo{journal}{arXiv: 1908.02038}} (\bibinfo{year}{2019}).


\bibitem{lammps}
\bibinfo{author}{Plimpton, S.}
\newblock \bibinfo{title}{Fast parallel algorithms for short-range molecular
 dynamics}.
\newblock \emph{\bibinfo{journal}{Journal of Computational Physics}}
 \textbf{\bibinfo{volume}{117}}, \bibinfo{pages}{1}
 (\bibinfo{year}{1995}).
%\newblock \urlprefix\url{http://lammps.sandia.gov}.

\bibitem{Wang2018}
\bibinfo{author}{Wang, Y.}, \bibinfo{author}{Hong, L.}, \bibinfo{author}{Wang,
 Y.}, \bibinfo{author}{Schirmacher, W.} \& \bibinfo{author}{Zhang, J.}
\newblock \bibinfo{title}{Disentangling boson peaks and van hove singularities
 in a model glass}.
\newblock \emph{\bibinfo{journal}{Phys. Rev. B}}
 \textbf{\bibinfo{volume}{98}}, \bibinfo{pages}{174207}
 (\bibinfo{year}{2018}).
  
\bibitem{Degiuli2011}
\bibinfo{author}{DeGiuli, E.} \& \bibinfo{author}{McElwaine, J.}
\newblock \bibinfo{title}{{Laws of granular solids: Geometry and topology}}.
\newblock \emph{\bibinfo{journal}{Physical Review E - Statistical, Nonlinear,
 and Soft Matter Physics}} \textbf{\bibinfo{volume}{84}}
 (\bibinfo{year}{2011}).

\bibitem{Lantagne-Hurtubise2017}
\bibinfo{author}{Lantagne-Hurtubise, {\'{E}}.}, \bibinfo{author}{Bhattacharjee,
 S.} \& \bibinfo{author}{Moessner, R.}
\newblock \bibinfo{title}{{Electric field control of emergent electrodynamics
 in quantum spin ice}}.
\newblock \emph{\bibinfo{journal}{Physical Review B}}
 \textbf{\bibinfo{volume}{96}}, \bibinfo{pages}{1--20} (\bibinfo{year}{2017}).
 
\bibitem{Xu2006}
\bibinfo{author}{Xu, C.}
\newblock \bibinfo{title}{{Gapless bosonic excitation without symmetry
 breaking: An algebraic spin liquid with soft gravitons}}.
\newblock \emph{\bibinfo{journal}{Physical Review B - Condensed Matter and
 Materials Physics}} \textbf{\bibinfo{volume}{74}}, \bibinfo{pages}{1--11}
 (\bibinfo{year}{2006}).
% 
 
%
\bibitem{deGiuli_PRE}
\bibinfo{author}{DeGiuli, E.}
\newblock \bibinfo{title}{{Edwards field theory for glasses and granular
 matter}}.
\newblock \emph{\bibinfo{journal}{Phys. Rev. E}} \textbf{\bibinfo{volume}{98}},
 \bibinfo{pages}{33001} (\bibinfo{year}{2018}).

\end{thebibliography}

%Supplemental Material for ``Emergent Elasticity in Amorphous Solids"
%%%%%%%%%%%%%%%%%%%%%%%%%%%%%%%%%%%%%%%%%%%%%%%%%%%%%%%%%
%%%%%%%%%%%%%%%%%%%%%%%%%%%%%%%%%%%%%%%%%%%%%%%%%%%%%%%%%

\clearpage

\begin{widetext}

\begin{appendix}

\section*{\large Supplemental Material for\\ ``Emergent Elasticity in Amorphous Solids"}

In this {supplementary document}, we describe in detail several key aspects of the theoretical framework and analysis of numerical and experimental data. In Section 1, we describe the methods used to generate the data used in this letter. In Section 2, we outline the derivation of the Gauss's law constraint on the Cauchy Stress tensor starting from the constraints of force and torque balance on every grain and discuss the mapping of grain-level properties  to the continuum theory.  In Section 3, we present results for the correlations of the electric displacement tensor $\hat{D}$, in a polarizable medium characterized by $\hat{\Lambda}$. Further in Section 4, we present experimental data for stress correlations from individual configurations. Finally, Section 5 describes the numerical results for the 2D system at finite temperature.

\subsection{Methods}

The main quantity of interest in this study, for a given packing is the stress tensor field in Fourier space given by
\begin{equation}
\hat{\sigma}^{p}\left({\bf q}\right)=\sum_{g=1}^{N_G^p} \hat{\sigma}_{g}^{p} \exp{\left(i{\bf q}\cdot{\bf r}_{g}^{\, p}\right)}.
\end{equation}

Here, `$p$' denotes a particular realization or packing of $N_G^p$ grains, while $g$ denotes a particular grain in the packing located at ${\bf r}_g^p$. $\hat{\sigma}_{g}^{p}$ represents the force moment tensor for the grain $g$, given by
\begin{equation}
\hat{\sigma}_{g}^{p}=\sum_{c=1}^{n_c^g} {\bf r}_c^{\, g} \otimes {\bf f}_c^g.
\end{equation} 
Here ${\bf r}_c^{\, g}$ denotes the position of the contact $c$ from the center of the grain $g$ and  ${\bf f}_c^g$ denotes the inter-particle force at the contact.

\subsubsection{Numerical Methods} 
We generate jammed packings of frictionless spheres interacting through one-sided spring potentials in two and three dimensions. Our implementation follows
the standard O'Hern protocol~\cite{ohern1,OHern_2003,Ramola2017}, with 
energy minimization  performed using two procedures (i) conjugate gradient minimization, and (ii) a FIRE~\cite{FIRE,FIRE2.0} minimization implementation in LAMMPS~\cite{lammps}. We have verified that these differences in protocol do not modify our results.

We simulate a 50:50 mixture of grains with diameter ratio 1:1.4. In our simulations, the system lengths are held fixed at $L_x = L_y = 1$ in 2D and $L_x = L_y = L_z = 1$ in 3D. We impose periodic boundary conditions in each direction, setting a lower cutoff between points in Fourier space $q_{min} = 2\pi$. We choose an upper cutoff $q_{max} = \pi/ d_{min}$ so as to not consider stress fluctuations occurring at length scales shorter than $d_{min}$, the diameter of the smallest grain in the packing. We have presented data for system sizes $N = 512,1024,2048,4096,8192$ in 2D, averaged over atleast $100$ configurations for each system size. The results obtained for different system sizes {have been} collapsed (see Fig. 1 of the main text) using the system size $N$ and $q_{max}$ as scaling parameters. This shows that the data presented is not significantly affected by finite size effects. All the 2D packings have {a} pressure per grain $P \in [0.016,0.017]$ and  packing fraction $\phi \in [0.878,0.882]$.  In 3D, the data for $N=27000$ is presented {in Fig. 2 of the main text, the data have been} averaged over $350$ configurations. The range of packing fractions for these configurations is $\phi \in [0.686,0.689]$, {with a} pressure per grain $P \in [0.0136,0.0147]$. 

\subsubsection{Experimental Methods}

The experimental {results were produced} from the analyses of {isotropically jammed packings and pure-sheared packings, which were both prepared} using a biaxial apparatus whose details can be found in Wang Et al. 2018~\cite{Wang2018}. This apparatus mainly consists of a rectangular frame mounted on top of a powder-lubricated horizontal glass plate. Each pair of parallel walls of the rectangular frame can move symmetrically with a motion precision of 0.1 mm so that the center of mass of the frame remains fixed. To apply isotropic compression, the two pairs of walls are programed to move inwards symmetrically. To apply pure shear, one pair of walls moves inwards, and the other pair of walls moves outwards, such that the area of the rectangle is kept fixed. The motion of walls is sufficiently slow to guarantee {that the deformation is quasi-static}. About $1.5$ m above the apparatus, there is an array of 2$\times$2 high-resolution (100 pixel/cm) cameras that are aligned and synchronized.

To prepare an isotropically jammed packing, we first filled the rectangular area with {a 50:50 mixture of 2680 bi-dispersed photoelastic disks (Vishay PSM-4), with diameters of 1.4 cm and 1.0 cm}, to create the various unjammed random initial configurations. Next, we applied isotropic compression to the disks to achieve a definite packing fraction $\phi$, which is the ratio between the area of disks and that of the rectangle.
To minimize the potential inhomogeneity  of force chains in the jammed packing, we constantly applied mechanical vibrations before the $\phi$ exceeded the jamming point $\phi_J\approx 84.0\%$ of frictionless particles.
The final isotropically jammed packing is confined in a square domain of 67.2 cm $\times$ 67.2 cm. Here, $\phi\approx 84.4\%$, the mean coordination number is around 4.1, the pressure is around 12 N/m, and {the} corresponding dimensionless pressure is $2\times 10^{-4}$.
Once the isotropically jammed packing was prepared, we then applied pure shear of strain $1.5\%$ to the packing to produce the pure-sheared packing. 
For both types of {packings}, two different images were recorded. Disk positions were obtained using the {\em normal image}, recorded without polarizers. Contact forces were analyzed from the {\em force-chain image}, recorded with polarizers, using the force-inverse algorithm~\cite{Trush_isotropic}.

\subsection{Mapping of Granular Media to Continuum Vector Charge Tensor Gauge Theory}

\begin{figure}[!htb]
\centering
\includegraphics[scale=0.5]{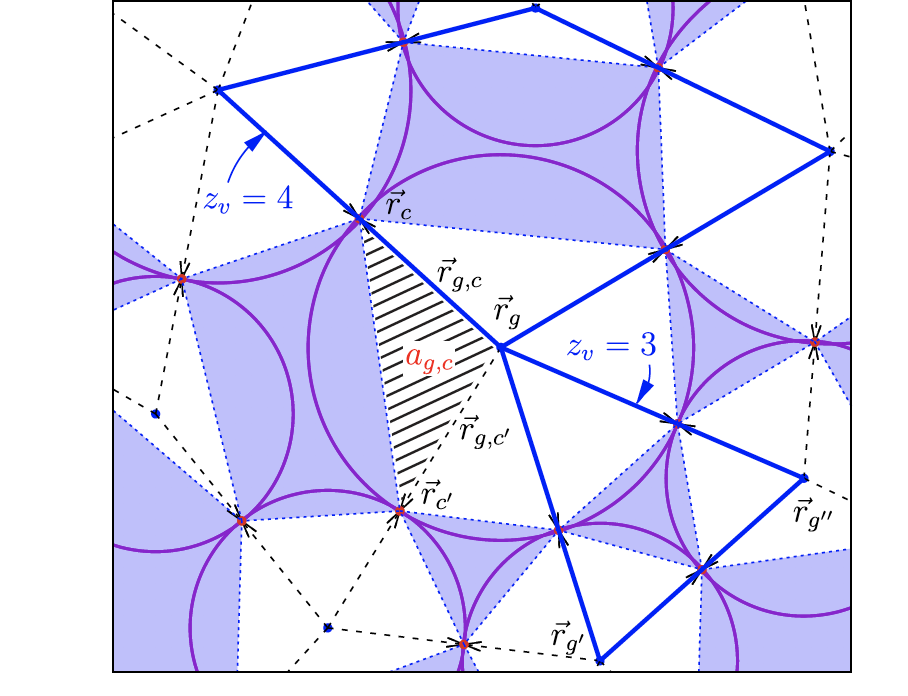}
\caption{ \small{A section of a jammed configuration of soft frictionless
disks in 2D. The centers of the grains  are located at
positions ${\bf r}_g$. The contact points between grains are located at
positions ${\bf r}_c$.  The triangle formed by the
points ${\bf r}_g,{\bf r}_{g^{\prime}},{\bf r}_c$ (shaded area) is uniquely assigned to the
contact c and has an associated area $a_{g,c}$.}}
\label{fig:network}
\end{figure}

The VCT Gauss's law (Eq. (2) in the main text),  is widely accepted as the coarse-grained description of stresses in athermal solids in mechanical equilibrium~\cite{Bouchaud-ideas,Lemaitre2018}.  Here, we  demonstrate the emergence of  this Gauss's law from local constraints of mechanical equilibrium, for the specific example of disordered granular solids.  The arguments can be easily generalized to other amorphous packings at zero temperature. 
Granular materials consist of an assembly of grains that interact with each other via contact forces, as shown in Fig. \ref{fig:network}.  In a granular solid, each grain is in mechanical equilibrium and thus, satisfy the constraints of force and torque balance.     The constraints of force and torque balance on a grain $g$,  with no ``body forces'' can be written as:
\begin{eqnarray}
\sum_{c \in g} {\bf f}_{g,c}  &=&0, \nonumber \\
\sum_{c \in g} {\bf r}_{g,c} \times {\bf f}_{g,c}  &= &0 ~,
\label{eq:forcebal}
\end{eqnarray}
respectively.   Here, ${\bf f}_{g,c} $ is the contact force,  and ${\bf r}_{g,c}$  the vector joining the center of grain $g$ to the contact $c$ (Fig. \ref{fig:network}).  This places $ dN+d(d-1)N$ nontrivial constraints on the $N$ grains that are part of the contact network. A grain is said to be a part of the contact network if it has more than one contact and grains which are not part of the contact network are defined to be ``rattlers''. In our representation, the rattlers become part of voids. Given a set of ${\bf f}_{g,c}$ and ${\bf r}_{g,c}$, one can define a stress tensor for  a  grain  with area $A_g$:
\begin{equation}
\hat {\sigma}_g = \frac{1}{A_g} \sum_{c \in g}  {\bf r}_{g,c} \otimes {\bf f}_{g,c} ~.
\label{eq:stress_micro}
\end{equation}
The coarse-grained stress tensor field, $\hat {D} ({\bf r})$ is obtained by summing $\hat {\sigma}_g$ over all grains included in a  coarse-graining volume, $\Omega_r$,  centered at ${\bf r}$:
\begin{equation}
\hat {D} ({\bf r}) = \frac{1}{\Omega_r} \sum_{g \in \Omega_r}  A_g \hat {\sigma}_g ~.
\label{eq:stress_coarse}
\end{equation}

The symmetry of $\hat {\sigma}_g$ is easy to establish by writing every contact force as the sum of a normal force, which is along the contact vector ${\bf r}_{g,c}$, and a tangential force perpendicular to it.   The normal part leads to a symmetric contribution to  $\hat {\sigma}_g$.   Using the torque-balance equation, Eq. \eqref{eq:forcebal},  the contribution from the tangential forces sum up to zero.   To establish the divergence free condition, we follow the approach outlined in Degiuli, E. and McElwaine, J. 2011~\cite{Degiuli2011}  by first subdividing $\hat{\sigma}_g$ into contributions from each contact.  As seen from Fig. \ref{fig:network}, we can associate a triangle of  area $a_{g,c}$ with each contact, and $A_g=\sum_{c \in g} a_{g,c}$.   Adopting a convention that we traverse around a grain in a  counterclockwise direction, we associate with contact $c$, the triangle that is defined by $c$ and  the contact $c^{\prime}$ that follows it.  We can then write: $A_g \hat {\sigma}_g  =  \sum_{c \in g} a_{g,c}  \hat{\sigma}_c$, where $ \hat{\sigma}_c$ is yet to be defined.   Comparing to Eq. \eqref{eq:stress_micro}, we see that $a_{g,c}  \hat{\sigma}_c = {\bf r}_{g,c} \otimes {\bf f}_{g,c} $, therefore $\hat{\sigma}_c ={\bf r}_{g,c} \otimes {\bf f}_{g,c} /a_{g,c}$.  The signed area $a_{g,c}$ is given by $a_{g,c}=(1/2)  {\bf r}_{g,c} \times ({\bf r}_{c^{\prime}} - {\bf r}_{c})$.  The divergence theorem is:  $\int_V \partial_i \sigma_{ij} =\int_{\partial V} n_i \sigma_{ij}$, where $\hat n$ is the unit normal to $\partial V$, which can be written as as $ \int_V \nabla \cdot  \hat {\sigma} = \int_{\partial V} (d{\bf r} \times \hat{\sigma})_j $.  We can apply the discrete version  of this  theorem to $\hat{\sigma}_g$ to get: 
\begin{equation}
A_g ( \nabla \cdot \hat{\sigma})_g = \sum_{c \in g} \hat{\sigma}_c \times {({\bf r}_{c^{\prime}} - {\bf r}_{c}) } = \sum_{c \in g} {\bf f}_{g,c} ={\bf f}_{ext}~.
\label{eq:stress_div}
\end{equation}
In the absence of external forces, $\hat{\sigma}_g$ is divergence free.   This grain-level condition leads to a similar condition on $\hat {D} ({\bf r})$: $\Omega_r \nabla \cdot \hat {D} ({\bf r}) = \sum_{c \in \partial \Omega}{\bf f}_c$, where the sum is over the contact forces on the boundary of $\Omega$, which is still discrete.

To map to the continuum theory, we posit that disorder averaging over all discrete networks that occur under given external conditions leads to

\begin{equation*}
\partial_i (\hat{D}({\bf r}))_{ij} =(f_{ext})_j.
\end{equation*} 

We expect this mapping to be accurate if the coarse-graining volume $\Omega$ is much larger than a typical grain volume. The excellent correspondence between disorder-averaged $\hat{D}$ correlations measured in granular packings and theoretical predictions, shown in the main text justifies the above mapping.  In Section 3 of this Supplementary Information, we present experimental measurements of $\hat{D}$ correlations in {\it individual} configurations to show that  self-averaging is a very good approximation for internal stresses in granular media.

\subsection{Stress-Stress Correlations in Polarizable Media}

In this section we present expressions for  the correlations of the $\hat{D}$ tensor, analogous to the expressions for the $\hat{E}$ correlations in vacuum (Eq. (7) in the main text). The starting point is Eq. (5) in the main text:  Gauss's law and the magnetostatic condition for a polarizable medium characterized by the rank-4 tensor, $\hat{\Lambda}$.  In the vacuum theory~\cite{Prem2018}, the strategy is to project out the divergence mode from the completely isotropic rank-4 tensor, using the magnetostatic condition.  This condition in ${\bf q}$- space, for a polarizable medium is given by
\begin{equation}
D_{ij}({\bf q})= (\hat{\Lambda}^{-1} \hat{A})_{ij} ({\bm q}) ;~~ A_{ij}({\bf q}) \equiv {\bf q} \otimes {\bm \phi} ~,
\label{eq:magneto_D}
\end{equation} 
where ${\bm \phi}$ is the electrostatic gauge potential, as discussed in the main text.  Since $\hat{\Lambda}$ has to obey the symmetry $ij \rightarrow ji$, it is simpler to write the components of $\hat{D}$ as a vector of length $3$ in 2D: $(D_{xx},~ D_{yy},~ D_{xy}$), and a vector of length $6$ in 3D.  The rank-4 tensor can be then expressed as a $3\times3$ (2D) and a $6 \times 6$ (3D) matrix~\cite{Otto2003}.  Furthermore, if $\hat{\Lambda}$ is a symmetric matrix in this representation,  then the $\hat{D}-\hat{D}$ correlations can be obtained from the $\hat{E}-\hat{E}$ correlations by a  transformation  of the metric: ${\bf q} \rightarrow {\bf \bar q} (\hat{\Lambda})$.  Such a transformation is reminiscent of the emergence of birefringence in quantum spin ice in the presence of an applied electric field~\cite{Lantagne-Hurtubise2017}. For the more general situation that can occur in granular media the matrix is not symmetric, and a cleaner approach is to use the dual formalism in which the  potential is obtained by solving Gauss's law~\cite{Xu2006}.  In this dual formalism,  potentials in 2D and 3D appear differently: a scalar in 2D and a second-rank symmetric tensor in 3D.  The expression for the correlations of the potentials can be worked out explicitly, and from that the $\hat{D}-\hat{D}$ correlations can be obtained in a straightforward manner.  
In 2D,  $\partial_i D_{ij} =0$ is solved by introducing a potential~\cite{Henkes2009,Lois2009,deGiuli_PRE,Xu2006,Ball_Blumenfeld}, $\psi$~:
$D_{ij} = \epsilon_{ia} \epsilon_{jb} \partial_a \partial_b \psi$.
The potential in 3D is a symmetric tensor, $\psi_{ij}$: $D_{ij} = \epsilon_{iab} \epsilon_{jcd} \partial_a \partial_c \psi_{bd}$

Here, we present the explicit construction of the correlations of $D_{ij}$ in 2D~\cite{Henkes2009,Lois2009,deGiuli_PRE}. The magnetostatic condition implies that $\hat{\Lambda}$ acts as a stiffness tensor in a Gaussian theory.  Using the ${\bm q}$-space representation: 
$D_{ij}({\bf q}) = \epsilon_{ia} \epsilon_{jb} q_a q_b \psi ({\bf q})$,
The correlations $\langle \psi (\bf{q}) \psi (-\bf{q}) \rangle$ can be computed, and give:
\begin{eqnarray}
\langle \psi (\bf{q}) \psi (-\bf{q}) \rangle &=& [A_{ij}({\bf{q}})  \Lambda_{ijkl} A_{kl} (-{\bf{q}}) ]^{-1}, \nonumber \\
A_{ij} &=&q^2 \delta_{ij} - q_i q_j.
\label{eq:corrpsi}
\end{eqnarray}
The correlations of $D_{ij}$ {then} follow as:
\begin{equation}
\langle D_{ij} ({\bf{q}}) D_{kl}(-{\bf{q}}) \rangle = \epsilon_{ia} \epsilon_{jb} \epsilon_{kc} \epsilon_{ld} q_a q_b q_c q_d \langle \psi (\bf{q}) \psi (-\bf{q}) \rangle. \nonumber \\
\label{eq:corrD}
\end{equation}

For the special case of $\hat{\Lambda}$ being a diagonal tensor with components $\lambda_{i}$, $i=xx~,yy~,xy$, the correlations simplify to:

\begin{align}
C_{xxxx}\left(\bf{q}\right)=\langle D_{xx}\left(\bf{q}\right)D_{xx}\left(-\bf{q}\right)\rangle & = \frac{q_y^4}{\lambda_{xx} q_y^4 + \lambda_{yy} q_x^4 + 2 \lambda_{xy}  q_x^2 q_y^2 },\nonumber \\
C_{xyxy}\left(\bf{q}\right)=\langle D_{xy}\left(\bf{q}\right)D_{xy}\left(-\bf{q}\right)\rangle & = \frac{q_x^2 q_y^2}{\lambda_{xx} q_y^4 + \lambda_{yy} q_x^4 + 2 \lambda_{xy}  q_x^2 q_y^2},  \nonumber \\
C_{yyyy}\left(\bf{q}\right)=\langle D_{yy}\left(\bf{q}\right)D_{yy}\left(-\bf{q}\right)\rangle & = \frac{q_x^4}{\lambda_{xx} q_y^4 + \lambda_{yy} q_x^4 + 2 \lambda_{xy}  q_x^2 q_y^2},  \label{eq:theory_original_corr_fn}\\
C_{xxxy}\left(\bf{q}\right)=\langle D_{xx}\left(\bf{q}\right)D_{xy}\left(-\bf{q}\right)\rangle & = -\frac{q_x q_y^3}{\lambda_{xx} q_y^4 + \lambda_{yy} q_x^4 + 2 \lambda_{xy}  q_x^2 q_y^2},  \nonumber \\
C_{xxyy}\left(\bf{q}\right)=\langle D_{xx}\left(\bf{q}\right)D_{yy}\left(-\bf{q}\right)\rangle & = \frac{q_x^2 q_y^2}{\lambda_{xx} q_y^4 + \lambda_{yy} q_x^4 + 2 \lambda_{xy}  q_x^2 q_y^2}, \nonumber \\
C_{xyyy}\left(\bf{q}\right)=\langle D_{xy}\left(\bf{q}\right)D_{yy}\left(-\bf{q}\right)\rangle & = -\frac{q_y q_x^3}{\lambda_{xx} q_y^4 + \lambda_{yy} q_x^4 + 2 \lambda_{xy}  q_x^2 q_y^2} . \nonumber
\end{align}

\begin{figure}
\centering
%\captionsetup{width=\linewidth}
\includegraphics[width=0.85\textwidth]{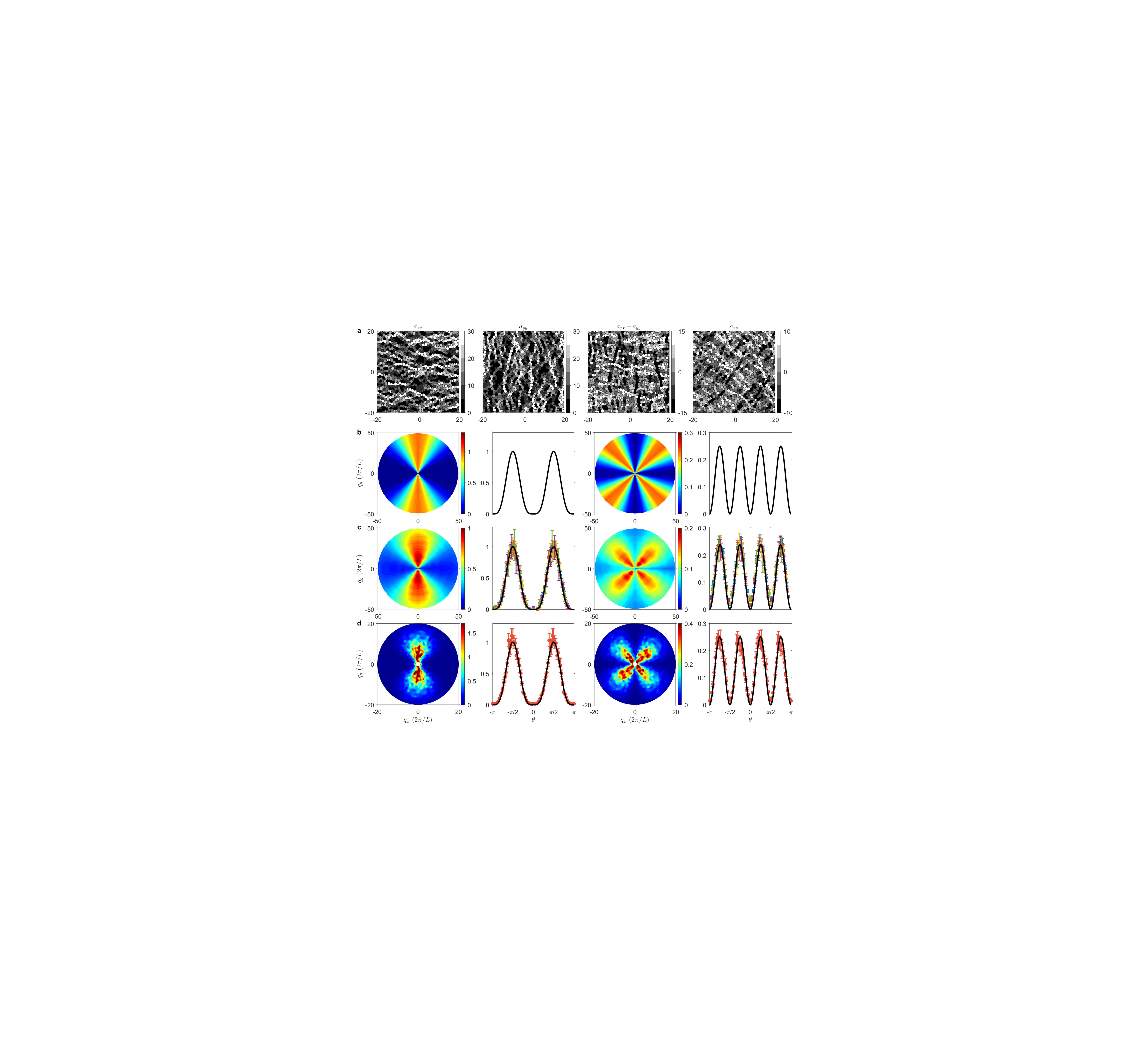}
\caption{{{\bfseries Comparisons in Fourier space between the theoretical predictions (black line) with $\Lambda = \mathbb{1}$, and the numerical and the experimental results (symbols) of the stress-stress correlations in 2D,  isotropically jammed systems.} {\bfseries a}, Photo-elastic images, in which each grain is shaded according to the magnitude of its normal stress, exhibit clear filamentary structures that are normally referred to as force chains. {\bfseries b}, Theoretical predictions of $C_{xxxx}(q,\theta)$ and $C_{xyxy}(q,\theta)$,  which are independent of $q$, and the corresponding angular functions $C_{xxxx} (\theta)$ and $C_{xyxy} (\theta)$. {\bfseries c}, Numerical data of the frictionless jammed packings within the range of pressure $P \in [0.016,0.017]$. The results of the five different system sizes $N = 512, 1024, 2048, 4096, 8192$ are shown in the angular plots. {\bfseries d}, Experimental data from {\it frictional} packings within the range of pressure $P \in [1.5\times10^{-4},2.9\times10^{-4}]$. \label{fig:2d_isotropic_suppl} All correlation functions are normalized by their peak values of $C_{xxxx}(\theta)$. The units of $q$ are $2\pi/L$, where $L$ is the system size:   $L\approx 100d_{\min}$ in simulations,  $L=40d_{\min}$ in experiments. Here $d_{\min}$ is the diameter of the small particle. Both the numerical and experimental data start to deviate from the theoretical predictions around $q \ge 2\pi/4d_{\min}$, indicating the breakdown of the continuum limit.}}

\label{fig:isotropic}
\end{figure}

The experimental and numerical measurements of correlations in 2D, shown in Fig. 1 of the main text and in Fig. \ref{fig:isotropic} of the supplementary, have been fit to the above forms. To analyze the correlations in isotropically compressed 3D packings, we assume that $\hat{\Lambda}$ is the identity tensor and use Eq. (7) of the main text, which gives the correlations in vacuum with an overall stiffness constant, $\lambda$.

\subsection{Force Chains and Stress Correlations}

 A  striking consequence of the anisotropic correlations in ${\bf q}$-space is evident if we analyze the correlations of the normal stresses, $D_{xx}$ and $D_{yy}$ in real space.   The Fourier Transform of  $C_{xxxx}$ in isotropic systems, with $\hat{\Lambda} = \lambda \mathbb{1}$ illustrates the point:
\begin{eqnarray}
C_{xxxx}(r_x,r_y) &= &\frac{3}{ 2 \lambda r_x ^2} ~~ {\rm for} ~ ~ r_x \gg r_y, \nonumber \\
C_{xxxx}(r_x,r_y) &= &-\frac{1}{2 \lambda  r_y^2} ~~ {\rm for} ~ ~ r_y \gg r_x.
\label{eq:forcechain}
\end{eqnarray}
The reverse is true for $C_{yyyy}$.  The consequence of this feature is that the  transverse correlations become {\it negatively} correlated.   The photo-elastic images from 2D experiments, shown in the main text and in Figs. \ref{fig:config1_suppl} and \ref{fig:config2_suppl}, are a striking visual representation of this stark difference between longitudinal and transverse correlations, which in turn is a manifestation of the conservation of ``charge-angular-momentum'', and the resulting sub-dimensional propagation~\cite{Pretko2017a}.  The $U(1)$ gauge theory with vector charges, therefore, clarifies the meaning  of force-chains within a continuum, disorder-averaged theory.

\subsubsection{Additional Analysis of Experiments}

In this subsection, we present results of stress correlations from individual configurations in the sheared experimental packings to illustrate how well self-averaging works in these jammed packings.  We note  that our systems are deep in the jammed region: we do not address the possible breakdown  of self-averaging close to the unjamming transition.
\begin{figure}[!htb]
\centering
\includegraphics[width=0.8\textwidth]{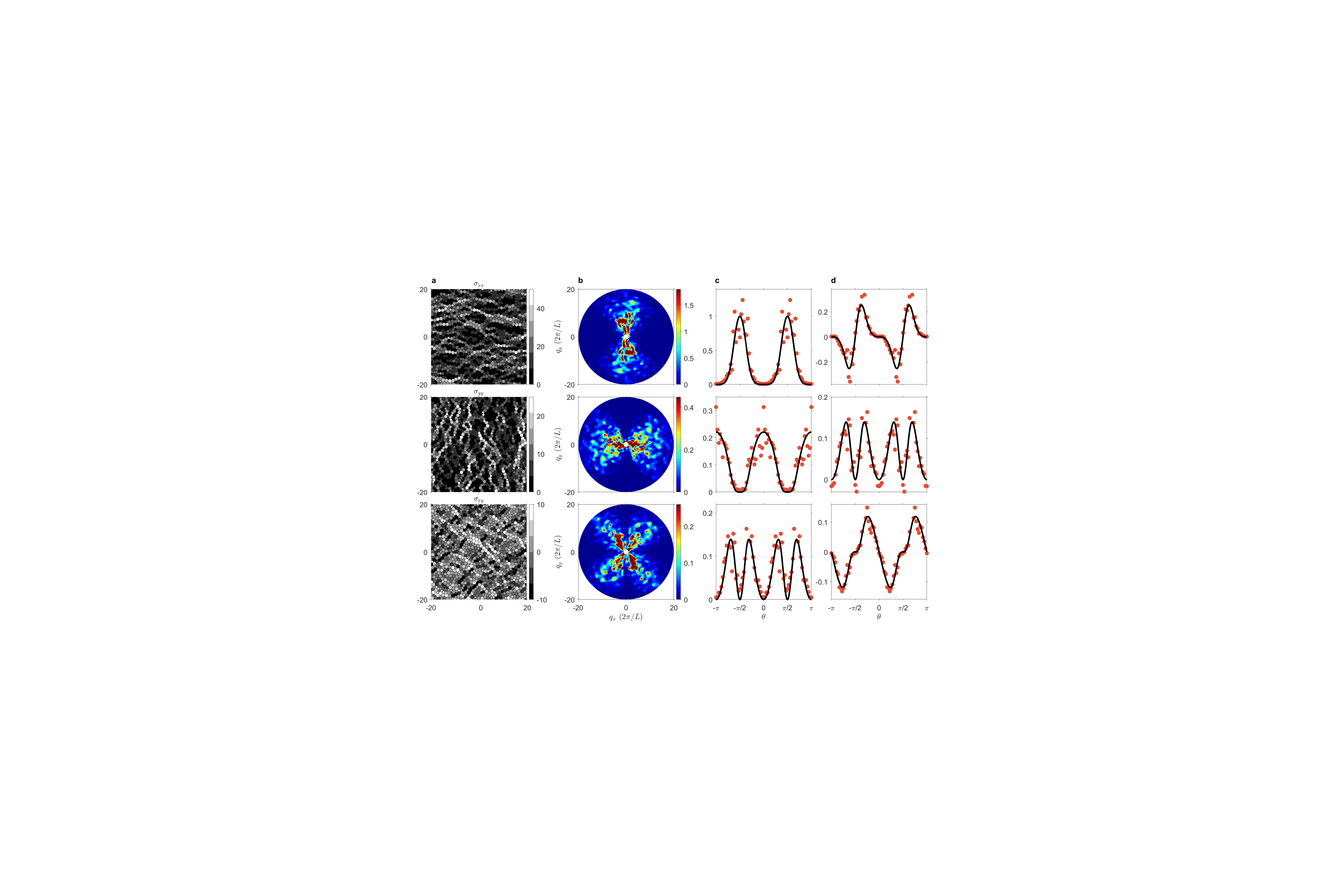}
\caption{Experimental  measurements of correlations in Fourier space, for a single packing in the ensemble of packings, used to generate the {{\it averaged}} correlations shown in the main text {(Fig. 3). The features observed in these} averaged correlations, are seen to emerge in a single packing, demonstrating the self-averaging property of the stress in these packings that are deep in the jammed regime. }
\label{fig:config1_suppl}
\end{figure}
\begin{figure}[!htb]
\centering
\includegraphics[width=0.8\textwidth]{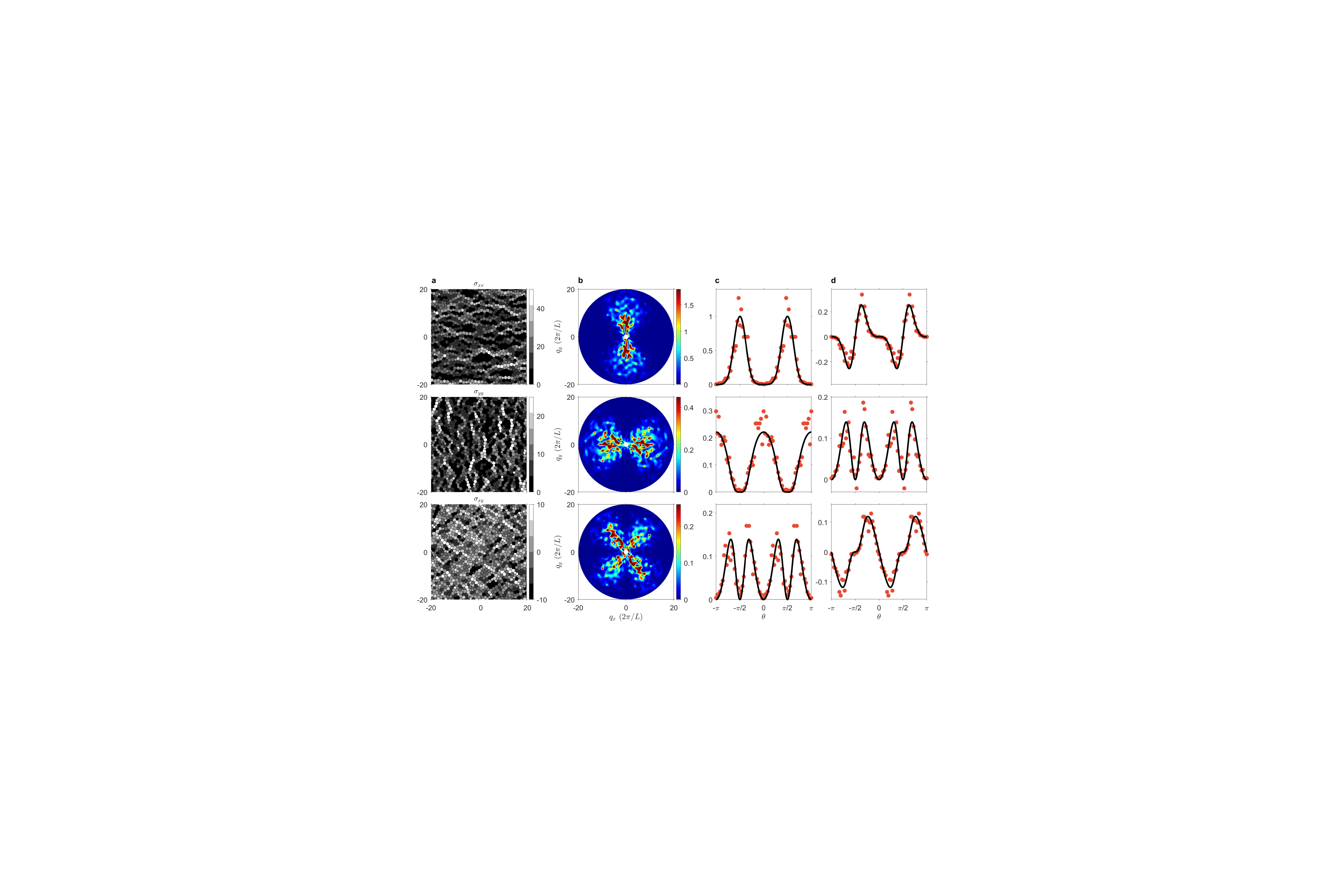}
\caption{ Experimental  measurements of correlations in Fourier space, for a second packing created under the same external conditions as in Fig. \ref{fig:config1_suppl}}
\label{fig:config2_suppl}
\end{figure}
\newpage

\FloatBarrier
\subsection{Finite Temperature Results}

\begin{figure}[!htb]

\includegraphics[width=0.9\textwidth]{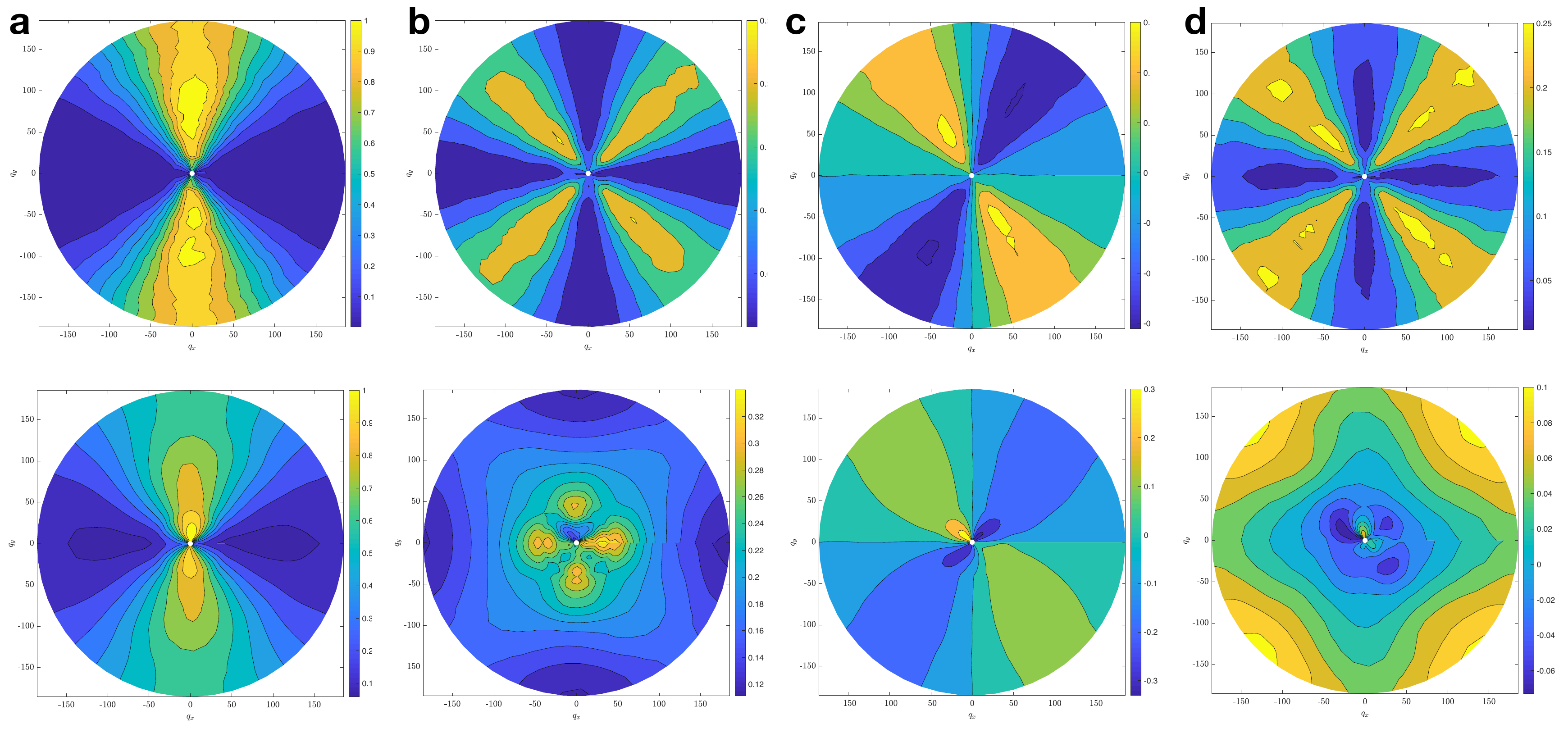}
\caption{Comparisons in Fourier space between stress correlations at zero ({\bfseries Top}) and finite ({\bfseries Bottom}) temperatures . The columns {\bfseries a}, {\bfseries b}, {\bfseries c}, and {\bfseries d} show the results for correlation functions $C_{xxxx}$, $C_{xyxy}$, $C_{xxxy}$ and $C_{xxyy}$ respectively. The packings used have an average compression energy per grain $E_{compression} \approx 10^{-4}$ and the finite temperature results have an average thermal energy per grain $E_{thermal} \approx 3.9 \times 10^{-4}$.}

\label{fig:finiteT}
\end{figure}

Pinch point singularities are one of {the salient} features of the VCT correlation functions. These singularities originate from the strict constraints of mechanical equilibrium imposed on athermal systems. For a system at finite temperature however, these constraints can be violated and hence we expect the pinch point singularities to disappear at finite {temperatures}. Thus, the presence of a pinch point singularity is a hallmark of an athermal system. The {numerically generated} stress correlations from a 2D system at finite temperature is shown in Fig. \ref{fig:finiteT} and it can be clearly seen that the pinch point singularity has vanished at this temperature $\left(E_{thermal}/E_{compression}=3.9\right)$.

The numerical simulations {were} carried out in LAMMPS and the finite temperature was imposed through a Nos\'e -Hoover thermostat. The protocol is to start with a valid athermal {$T=0$} configuration, generated following the procedure described in the Numerical Methods Section and then perform finite temperature dynamics to compute the stress correlations at {a non-zero} temperature. This procedure is then repeated over multiple initial athermal configurations and {ensemble averaged} to obtain the finite temperature stress correlations. The results {displayed} are obtained for packings of $8192$ disks with an average pressure per grain $P \in [0.016,0.017]$. The results shown {have been} averaged over 50 starting athermal configurations in 2D with 50 finite temperature {configurations sampled during the finite temperature molecular dynamics run, for each of the 50 starting configurations}.

\end{appendix}

\clearpage

\end{widetext}

\end{document}